\documentclass[10pt,letterpaper,american,showpacs,prb,twocolumn,aps,superscriptaddress]{revtex4}
\usepackage[T1]{fontenc}
\usepackage[latin1]{inputenc}
\usepackage{bm}
\usepackage{amsmath}
\usepackage{babel}
\usepackage{graphics}
\usepackage{amssymb}
\usepackage{hyperref}
\makeatletter

\usepackage{pifont}

\makeatother
\begin{document}

\title{Non-Markovian dynamics in a spin star system: Exact solution and
approximation techniques}

\author{Heinz-Peter Breuer}
\affiliation{Physikalisches Institut, Universität Freiburg,
Hermann-Herder-Str.~3, D-79104 Freiburg, Germany}

\author{Daniel Burgarth}
\affiliation{Physikalisches Institut, Universität Freiburg,
Hermann-Herder-Str.~3, D-79104 Freiburg, Germany}

\author{Francesco Petruccione}
\affiliation{Physikalisches Institut, Universität Freiburg,
Hermann-Herder-Str.~3, D-79104 Freiburg, Germany}
\affiliation{School of Pure and Applied Physics, University of KwaZulu-Natal, 4041 Durban, South Africa}

\begin{abstract}
The reduced dynamics of a central spin coupled to a bath of
$N$ spin-$\frac{1}{2}$ particles arranged in a spin star
configuration is investigated. The exact time evolution of the
reduced density operator is derived, and an analytical solution is
obtained in the limit $N\rightarrow\infty$ of an infinite number
of bath spins, where the model shows complete relaxation and
partial decoherence. It is demonstrated that the dynamics of the
central spin cannot be treated within the Born-Markov
approximation. The Nakajima-Zwanzig and the time-convolutionless
projection operator technique are applied to the spin star system.
The performance of the corresponding perturbation expansions of
the non-Markovian equations of motion is examined through a
comparison with the exact solution.
\end{abstract}
\maketitle

\section{Introduction}

\newcommand{\bra}[1]{\left\langle #1 \right| }

\newcommand{\ket}[1]{\left| #1 \right\rangle }

Solid state spin nanodevices are known as very promising
candidates for quantum computation \cite{MNIC,S29,BK98,DDA02} and also
for quantum communication \cite{SB03}. They provide a scalable
system that can easily be integrated into standard silicon
technology. A drawback of such systems compared to other
proposals for quantum computing, such as ion traps \cite{JCPZ} and
cavity QED \cite{TPSG+95}, are the many degrees of freedom of the
surrounding material causing dissipation and decoherence
\cite{HBFP02b}. The first step in overcoming these disadvantages
is to be able to model them.

An important contribution to quantum noise in solid state
systems arises from the nuclear spins, and recently much work has
been devoted to the modeling of spin bath systems
\cite{AKDL,DM02,RdS03,S23,S7,S8,S9,S10,S3} (for a review, see
Refs. \onlinecite{JSAK03,S2}). The interaction of a central spin with a bath
of environmental spins often leads to strong non-Markovian
behavior. The usual derivations of Markovian quantum master
equations known, e.~g., from atomic physics \cite{CCJD98} and
quantum optics \cite{CGPZ00} therefore fail for many spin bath
models and a detailed investigation of methods is
required which are capable of going beyond the Markovian
approximation.

In this paper, we examine the reduced dynamics of a simple
spin star system. The advantage of this model is that, while
showing several interesting features such as partial decoherence
and strong non-Markovian behavior, it is exactly solvable due to
its high symmetry. The model therefore represents an appropriate
example for a general discussion of the performance of various
non-Markovian methods. We study here the Nakajima-Zwanzig \cite{SN58,RZ60,HM65} and the
time-convolutionless projection operator technique \cite{FSTA80,AR72,NvK74,vK74}
and derive and analyze the perturbation expansions of the
corresponding non-Markovian master equations.

The paper is organized as follows. In Sec.~\ref{exact}, we
introduce the model investigated, a spin star model involving a
Heisenberg XX coupling (Sec.~\ref{model}), and determine the exact
time evolution of the central spin (Sec.~\ref{reduceddyn}). In
Sec.~\ref{limit} we analyze the limit of an infinite number of
bath spins, discuss the behavior of the von Neumann entropy of
the central spin, and demonstrate that the model exhibits complete
relaxation and partial decoherence. Several non-Markovian
approximation techniques are discussed in
Sec.~\ref{approximations}. The dynamic equations found in second
order in the coupling are introduced in Sec.~\ref{second}, where
it is also demonstrated that the prominent Born-Markov
approximation is not applicable to the spin star model. Employing
a technique which enables the calculation of the correlation
functions of the spin bath, we derive in Sec.~\ref{higher} the
perturbation expansions corresponding to the Nakajima-Zwanzig and
to the time-convolutionless projection operator technique and
compare these approximations with the analytical solution for the
dynamics of the central spin. Finally, the conclusions are drawn
in Sec.~\ref{CONCLU}.

\section{\label{exact}Exact dynamics}

\subsection{The model\label{model}}

We consider a ``spin star'' configuration \cite{AHSB} which
consists of \( N+1 \) localized spin-\( \frac{1}{2} \)
particles. One of the spins is located at the center of the star,
while the other spins, labeled by an index \( i=1,2,\ldots ,N
\), surround the central spin at equal distances on a sphere. In
the language of open quantum systems \cite{HBFP02b} we regard the
central spin with Pauli spin operator $\bm{\sigma}$ as an
open system living in a two-dimensional Hilbert space \(
\mathcal{H}_{S} \) and the surrounding spins described by the spin
operators $\bm{\sigma}^{(i)}$ as a spin bath with Hilbert space
\( \mathcal{H}_{B} \) which is given by an $N$-fold tensor product
of two dimensional spaces.

The central spin $\bm{\sigma}$ interacts with the bath spins
$\bm{\sigma}^{(i)}$ via a Heisenberg XX interaction \cite{ELTS61}
represented through the Hamiltonian
\begin{equation} \label{hamiltonian}
 \alpha H=2\alpha \left( \sigma_{+}J_{-}+\sigma _{-}J_{+}\right),
\end{equation}
where
\begin{equation}
 J_{\pm } \equiv \sum ^{N}_{i=1}\sigma ^{(i)}_{\pm },
\end{equation}
and
\begin{equation}
 \sigma ^{(i)}_{\pm } \equiv
 \frac{1}{2}\left( \sigma ^{(i)}_{1}\pm i\sigma ^{(i)}_{2}\right)
\end{equation}
represents the raising and lowering operators of the $i$th
bath spin. The Heisenberg XX coupling has been found to be an
effective Hamiltonian for the interaction of some quantum dot
systems \cite{AIDA+99}. Equation (\ref{hamiltonian}) describes a
very simple time independent interaction with equal coupling
strength \( \alpha  \) for all bath spins. It is invariant under
rotations around the \( z \)-axis. The operator \( \mathbf{J}\equiv
\frac{1}{2}\sum ^{N}_{i=1}\bm{\sigma }^{(i)} \) represents the
total spin angular momentum of the bath (units are chosen
such that $\hbar = 1$). The central spin thus couples to the
collective bath angular momentum.

We introduce an ON basis in the bath Hilbert space \(
\mathcal{H}_{B} \) consisting of states \( \ket{j,m,\nu } \).
These states are defined as eigenstates of \( J_{3} \) (eigenvalue
\( m \)) and of \( \mathbf{J}^{2} \) [eigenvalue \( j(j+1) \) ]. The
index \( \nu  \) labels the different eigenstates in the
eigenspace \( \mathcal{M}_{j,m} \) belonging to a given pair \(
(j,m) \) of quantum numbers. As usual, \( j\leq \frac{N}{2} \) and
\( -j\leq m\leq j \). The dimension of \( \mathcal{M}_{j,m} \) is given by the
expression \cite{JWKM,AHSB}
\begin{equation}
 n(j,N) = \left( \begin{array}{c} N \\ N/2-j \end{array} \right)
 -\left( \begin{array}{c} N \\ N/2-j-1 \end{array} \right).
\end{equation}

We further introduce the usual superoperator notation for the Liouville
operator
\begin{equation}
 \mathcal{L}\rho (t)\equiv -i\left[ H,\rho (t)\right] ,
\end{equation}
where \( \rho (t) \) denotes the density matrix of the total
system in the Hilbert space \( \mathcal{H}_{S}\otimes
\mathcal{H}_{B} \). The formal solution of the von Neumann
equation
\begin{equation}
 \frac{d}{dt}\rho (t)= \alpha \mathcal{L}\rho (t)
\end{equation}
can then be written as
\begin{equation} \label{full_dynamics}
 \rho (t)=\exp \left( \alpha \mathcal{L}t\right) \rho (0).
\end{equation}
Our main goal is to derive the dynamics of the reduced
density matrix
\begin{equation} \label{reduce_definition}
 \rho_{S}(t)\equiv \textrm{tr}_{B}\left\{ \rho (t)\right\},
\end{equation}
where \( \textrm{tr}_{B} \) denotes the partial trace taken
over the Hilbert space \( \mathcal{H}_{B} \) of the spin bath. The
reduced density matrix is completely determined in terms of the
Bloch vector
\begin{equation}
 \mathbf{v}(t)=\left(
 \begin{array}{c} v_{1}(t)\\ v_{2}(t)\\ v_{3}(t)
 \end{array}\right) \equiv \textrm{tr}_{S}\left\{ \bm{\sigma }\rho _{S}(t)\right\}
\end{equation}
through the relationship
\begin{eqnarray}
 \rho _{S}(t)= & \frac{1}{2}\left( \begin{array}{cc}
 1+v_{3}(t) & v_{1}(t)-iv_{2}(t)\\
 v_{1}(t)+iv_{2}(t) & 1-v_{3}(t)
 \end{array}\right), \label{relationship_bloch_dens}
\end{eqnarray}
where \( \textrm{tr}_{S} \) denotes the partial trace over
the Hilbert space \( \mathcal{H}_{S} \) of the central spin. We
note that the length \( r(t)\equiv \left| \mathbf{v}(t)\right| \) of
the Bloch vector is equal to $1$ if and only if \( \rho _{S}(t) \)
describes a pure state, and that the von Neumann entropy $S$ of
the central spin can be expressed as a function of the length \(
r(t) \) of the Bloch vector:
\begin{eqnarray}
 S &\equiv& \textrm{tr}_{S}\left\{ -\rho_{S}\ln \rho _{S}\right\} \\
 &=& \ln 2 - \frac{1}{2}(1-r)\ln (1-r) + \frac{1}{2}(1+r)\ln (1+r).
 \nonumber
\end{eqnarray}

The initial state of the reduced system at \( t=0 \) is
taken to be an arbitrary (possibly mixed) state
\begin{eqnarray} \rho _{S}(0)= & \left( \begin{array}{cc}
\frac{1+v_{3}(0)}{2} & v_{-}(0)\\ v_{+}(0) & \frac{1-v_{3}(0)}{2}
\end{array}\right),
\end{eqnarray}
while the spin bath is assumed to be in an unpolarized
infinite temperature state:
\begin{equation}
 \rho _{B}(0) = 2^{-N} I_{B}.
\end{equation}
Here, $I_B$ denotes the unit matrix in ${\mathcal{H}}_B$, and
we have defined the \( v_{\pm } \) as linear combinations of the
components \( v_{1,2} \) of the Bloch vector:
\begin{equation}
 v_{\pm }=\frac{v_{1}\pm iv_{2}}{2}.
\end{equation}
The initial state of the total system is given by an
uncorrelated product state \( \rho _{S}(0)\otimes \rho
_{B}(0).\)

\subsection{Reduced System Dynamics\label{reduceddyn}}

In this section, we will derive the exact dynamics of the
reduced density matrix \( \rho _{S}(t) \) for the model given
above. One possibility of obtaining the evolution of the central
spin is to substitute Eq.~(\ref{full_dynamics}) into
Eq.~(\ref{reduce_definition}) and to expand the exponential with
respect to the coupling. This yields
\begin{eqnarray}
 \rho _{S}(t) &\equiv& \textrm{tr}_{B}\left\{
 \exp \left( \alpha \mathcal{L}t\right) \rho _{S}(0)\otimes \rho _{B}(0)\right\}
 \nonumber \\
 &=& \sum^{\infty }_{k=0}\frac{(\alpha t)^k}{k!}
 \textrm{tr}_{B}\left\{
 \mathcal{L}^{k}\rho _{S}(0)\otimes 2^{-N}I_{B}\right\}.
 \label{reduced_definit}
\end{eqnarray}
It is easy to verify that we have
\begin{equation}
 H^{2n} = 4^n \left[ \sigma_+\sigma_-(J_-J_+)^n
 +\sigma_-\sigma_+(J_+J_-)^n \right]
\end{equation}
and
\begin{equation}
 H^{2n+1} = 2\cdot 4^n \left[ \sigma_+\sigma_-(J_+J_-)^n
 +\sigma_-\sigma_+(J_-J_+)^n \right].
\end{equation}
We note that such simple expressions are obtained since a term
$\sigma_3J_3$ is missing in the interaction Hamiltonian. We
substitute the last two equations into
\begin{equation}
 {\mathcal{L}}^n \rho = i^n\sum_{k=0}^n (-1)^k
 \binom{n}{k} H^k \rho H^{n-k}
\end{equation}
to get the formulas
\begin{equation} \label{L^2m+1_rho}
 \textrm{tr}_{B}\left\{ \mathcal{L}^{2k-1}\rho _{S}(0)\otimes 2^{-N} I_{B}\right\} =0
\end{equation}
and
\begin{eqnarray}
 \lefteqn {\textrm{tr}_{B}\left\{ \mathcal{L}^{2k}\rho _{S}(0)
 \otimes 2^{-N} I_{B} \right\}
 = \left( -16\right) ^{k}Q_{k}\frac{v_{3}(0)}{2}\sigma _{3}}
 && \label{L^2m_rho} \\
 &+& \left( -4\right) ^{k}\left( \sum ^{k}_{l=0}
 \left( \begin{array}{c} 2k \\ 2l \end{array} \right)
 R^{k-l}_{l}\right) \left( v_{-}(0)\sigma _{+}+v_{+}(0)\sigma _{-}\right),
 \nonumber
\end{eqnarray}
which hold for all $k=1,2,\ldots$. Here, we have introduced
the bath correlation functions
\begin{eqnarray}
 Q_{k} &\equiv& \frac{1}{2^{N}}\textrm{tr}_{B}\left\{\left( J_{+}J_{-}\right)^{k}
 \right\}, \\
 R^{k-l}_{l} &\equiv&
 \frac{1}{2^{N}}\textrm{tr}_{B}\left\{ \left( J_{+}J_{-}\right)^{k-l}
 \left( J_{-}J_{+}\right) ^{l}\right\} .
\end{eqnarray}
We will come back to these correlation functions when we
discuss approximation techniques in Sec.~\ref{approximations}.

Using the formulas (\ref{L^2m+1_rho}) and (\ref{L^2m_rho}) in
Eq.~(\ref{reduced_definit}) we can express the components of the
Bloch vector as follows,
\begin{eqnarray}
 v_{\pm }(t) &=& f_{12}(t)v_{\pm}(0), \label{bloch1} \\
 v_{3}(t) &=& f_{3}(t)v_{3}(0),\label{bloch2}
\end{eqnarray}
where we have introduced the functions
\begin{eqnarray}
 \lefteqn {f_{12}(t)\equiv } &  & \\
 && {\mathrm{tr}}_B \left\{
 \cos\left[ 2\sqrt{J_+J_-}\alpha t\right]
 \cos\left[ 2\sqrt{J_-J_+}\alpha t\right] \otimes 2^{-N}I_B
 \right\}, \nonumber
\end{eqnarray}
and
\begin{equation}
 f_{3}(t) \equiv {\mathrm{tr}}_B \left\{
 \cos\left[ 4\sqrt{J_+J_-}\alpha t\right] \otimes 2^{-N}I_B
 \right\}.
\end{equation}
Calculating the traces over the spin bath in the eigenbasis of \(
J_{3} \) and \( \mathbf{J}^{2} \) using
\begin{equation}
 J_{\mp }J_{\pm }\ket{j,m,\nu }=\left( j\mp m\right)
 \left( j\pm m+1\right) \ket{j,m,\nu},
\end{equation}
we find
\begin{eqnarray} \label{F12}
 \lefteqn {f_{12}(t)\equiv } &  & \nonumber \\
 && \sum _{j,m}n(j,N)\frac{\cos \left[ 2h(j,m)\alpha t\right]
 \cos \left[ 2h(j,-m)\alpha t\right] }{2^{N}}, \qquad
\end{eqnarray}
and
\begin{equation} \label{F3}
 f_{3}(t) \equiv \sum _{j,m}n(j,N)\frac{\cos
 \left[ 4h(j,m)\alpha t\right]}{2^{N}},
\end{equation}
where we have introduced the quantity \( h(j,m)\equiv
\sqrt{(j+m)(j-m+1)} \).

Thus we have determined the exact dynamics of the reduced
system: The density matrix $\rho_S(t)$ of the central spin is
given through the components of the Bloch vector which are
provided by the relations (\ref{bloch1}), (\ref{bloch2}) and (\ref{F12}), (\ref{F3}).
We note that
the dynamics can be expressed completely through only two
real-valued functions \( f_{12}(t) \) and \( f_{3}(t) \). This
fact is connected to the rotational symmetry of the system.

The reduced system dynamics has been obtained above with the
help of an expansion of $\exp(\alpha{\mathcal{L}}t)$ with respect
to the coupling constant $\alpha$. It should be clear that,
alternatively, the behavior of the central spin may be found
directly from the solution of the Schr\"odinger equation for the
composite system. This solution can easily be constructed by
making use of the fact that the subspaces spanned by the states
$|+\rangle\otimes|j,m,\nu\rangle$ and
$|-\rangle\otimes|j,m+1,\nu\rangle$ are invariant under the time
evolution, where $|\pm\rangle$ denotes the eigenstate of
$\sigma_3$ belonging to the eigenvalue $\pm 1$.

Sometimes it is useful to express the reduced dynamics in terms of
superoperators instead of the Bloch vector. To this end, we
introduce superoperators $\mathcal{S}_{\pm}$ and
$\mathcal{S}_{3}$, which are defined by their action on an
arbitrary operator \( A \):
\begin{eqnarray}
 \mathcal{S}_{\pm }A & \equiv &
 \sigma _{\pm }A\sigma _{\mp }-\frac{1}{2}\left\{
 \sigma_{\mp }\sigma _{\pm },A\right\} ,\label{supops} \\
 \mathcal{S}_{3}A & \equiv  & \sigma _{3}A\sigma_{3} - A. \label{supops2}
\end{eqnarray}
With these definitions we may write the reduced density matrix as
follows,
\begin{eqnarray}
 \lefteqn {\rho _{S}(t)=\frac{I_{S}}{2}} &  &
 \label{supop_rho_s_von_t} \\
 && +\frac{1}{2}\left[
 \left(\frac{1}{2}f_{3}(t)-f_{12}(t)\right)\mathcal{S}_{3}
 -f_{3}(t)\left( \mathcal{S}_{+}
 +\mathcal{S}_{-}\right) \right] \rho _{S}(0), \nonumber
\end{eqnarray}
where $I_S$ denotes the unit matrix in ${\mathcal{H}}_S$.
Due to the non-unitary behavior of the reduced system, the
superoperator on the right-hand side is not invertible for
all times. This point will become important later on when
we discuss approximation strategies.

\subsection{\label{limit}The Limit of an infinite number of bath spins}

\begin{figure}[htba]
{\centering
\resizebox*{0.47\textwidth}{!}{\includegraphics{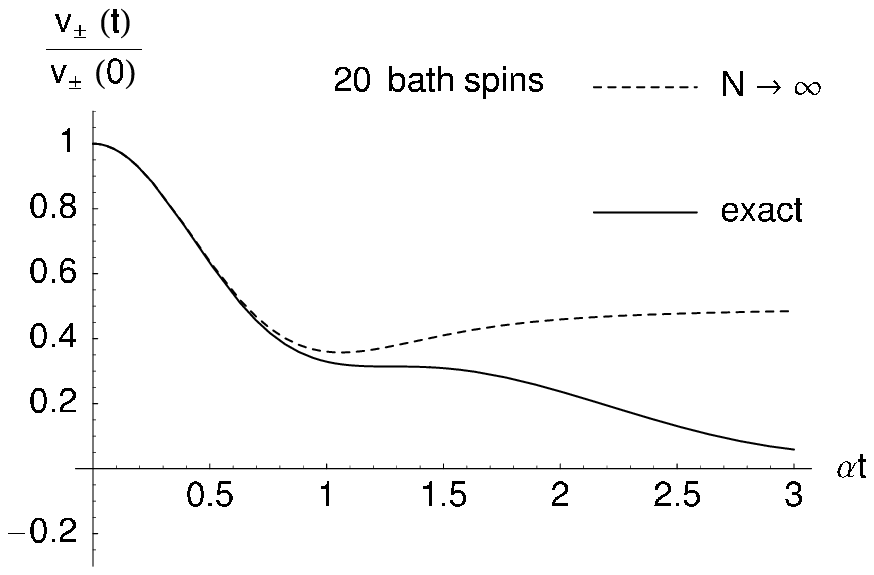}}
\resizebox*{0.47\textwidth}{!}{\includegraphics{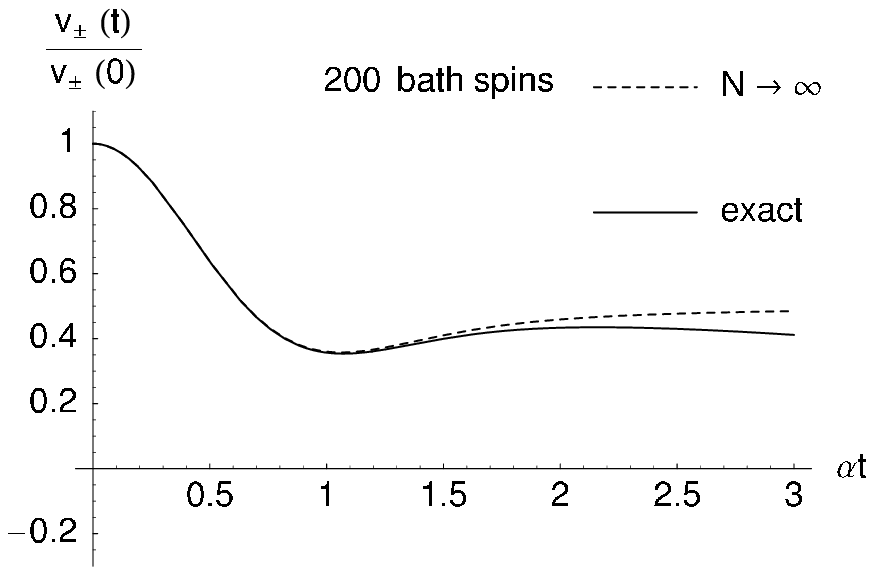}} \par}

\caption{\label{ninf1}Comparison of the limit \protect\(
N\rightarrow \infty \protect \) [see Eqs.~(\ref{F123infty}) and
(\ref{Gtinfty})] with the exact functions for \protect\(
N=20\protect \) and \protect\( N=200.\protect \) The plot shows
the \protect\( v_{\pm }\protect \)-component of the Bloch
vector.}
\end{figure}

\begin{figure}[htba]
{\centering \resizebox*{0.47\textwidth}{!}{\includegraphics{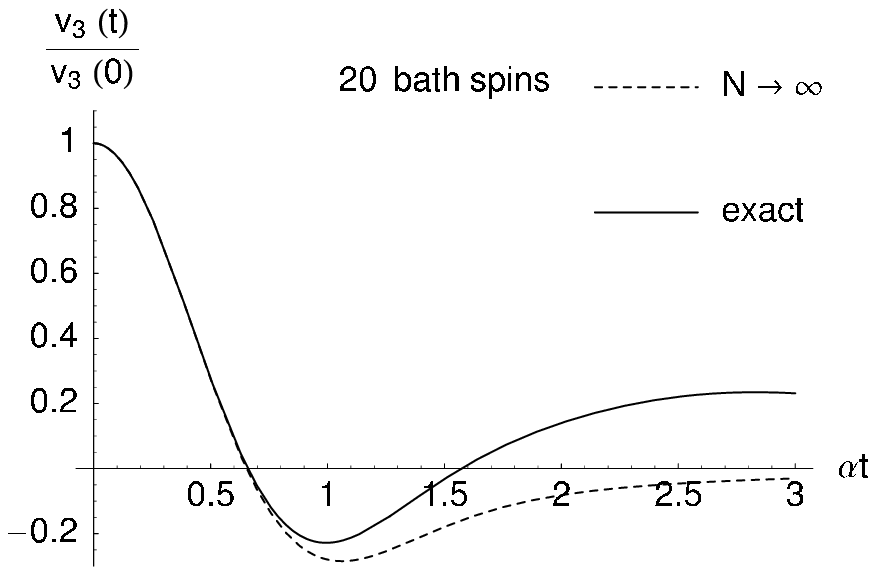}}
\resizebox*{0.47\textwidth}{!}{\includegraphics{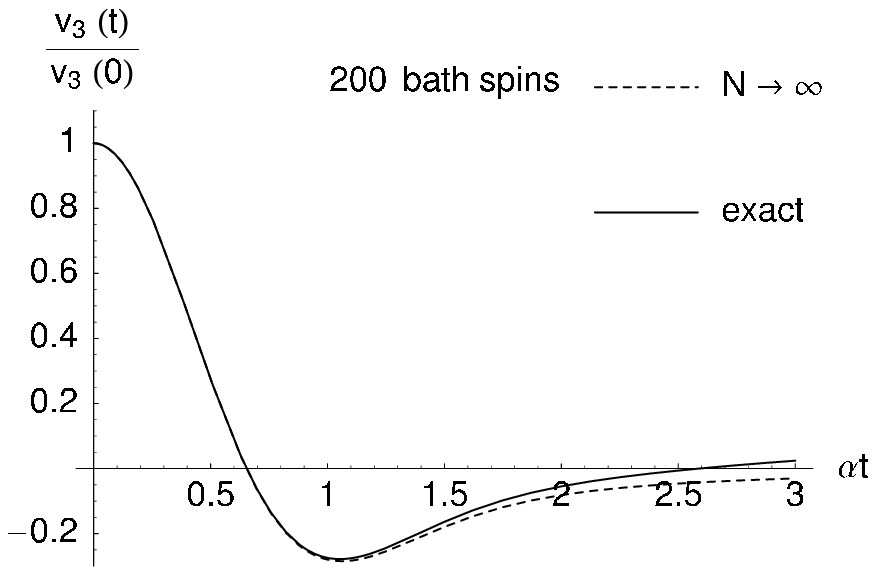}} \par}

\caption{\label{ninf2}Comparison of the limit \protect\(
N\rightarrow \infty \protect \) [see Eqs.~(\ref{F123infty}) and
(\ref{Gtinfty})] with the exact functions for \protect\(
N=20\protect \) and \protect\( N=200.\protect \) The figure shows
the \protect\( v_{3}\protect \)-component of the Bloch vector.}
\end{figure}

The explicit solution constructed in the previous section takes on
a relatively simple form in the limit \( N\rightarrow \infty  \)
of an infinite number of bath spins. It is demonstrated in the
appendix that for large \( N \) the bath correlation functions
approach the asymptotic expression
\begin{equation} \label{ninfty2}
 Q_{k}\approx R^{k-l}_{l}\approx \frac{k!}{2^{k}}N^{k}.
\end{equation}
Consequently, in Eq.~(\ref{reduced_definit}) a term of the order
$N^k$ always occurs together with a factor of $\alpha^{2k}$. A
non-trivial finite limit $N\rightarrow\infty$ therefore exists if
we rescale the coupling constant as follows,
\begin{equation}
 \alpha \rightarrow \frac{\alpha }{\sqrt{N}}.
\end{equation}
Using this approximation in Eq.~(\ref{L^2m_rho}) one obtains
\begin{eqnarray}
 \lefteqn {\textrm{tr}_{B}\left\{
 \mathcal{L}^{2k}\rho _{S}(0)\otimes 2^{-N}I_B\right\} }
 && \nonumber \\
 && \approx \frac{\left( -8N\right)^{k}k!}{2}
 \left( v_{3}(0)\sigma _{3}+v_{-}(0)\sigma _{+}+v_{+}(0)\sigma _{-}\right).
 \quad
\end{eqnarray}
If we insert this into Eq.~(\ref{reduced_definit}) we get an
infinite series which yields the following expressions for the
functions $f_{12}(t)$ and $f_3(t)$,
\begin{equation} \label{F123infty}
 f_{12}(t) =  1+g(t), \qquad f_{3}(t) = 1+2g(t),
\end{equation}
where
\begin{equation} \label{Gtinfty}
 g(t)\equiv -\alpha t\exp (-2\alpha ^{2}t^{2})\sqrt{\frac{\pi }{2}}
 \textrm{erfi}(\sqrt{2}\alpha t).
\end{equation}
Note that \( \textrm{erfi}(x) \) is the imaginary error function.
It is a real-valued function defined by
\begin{equation}
 \textrm{erfi}(x)\equiv \frac{\textrm{erf}(ix)}{i},
\end{equation}
which leads to the Taylor series
\begin{equation}
 \textrm{erfi}(x) =
 \frac{2}{\sqrt{\pi }}\sum ^{\infty}_{k=0}\frac{x^{2k+1}}{k!(2k+1)}.
\end{equation}
Figures \ref{ninf1} and \ref{ninf2} show that this
approximation obtained in the limit of an infinite number of bath
spins is already reasonable for \( N\approx 200 \).

\begin{figure}[htba]
{\centering \resizebox*{0.47\textwidth}{!}{\includegraphics{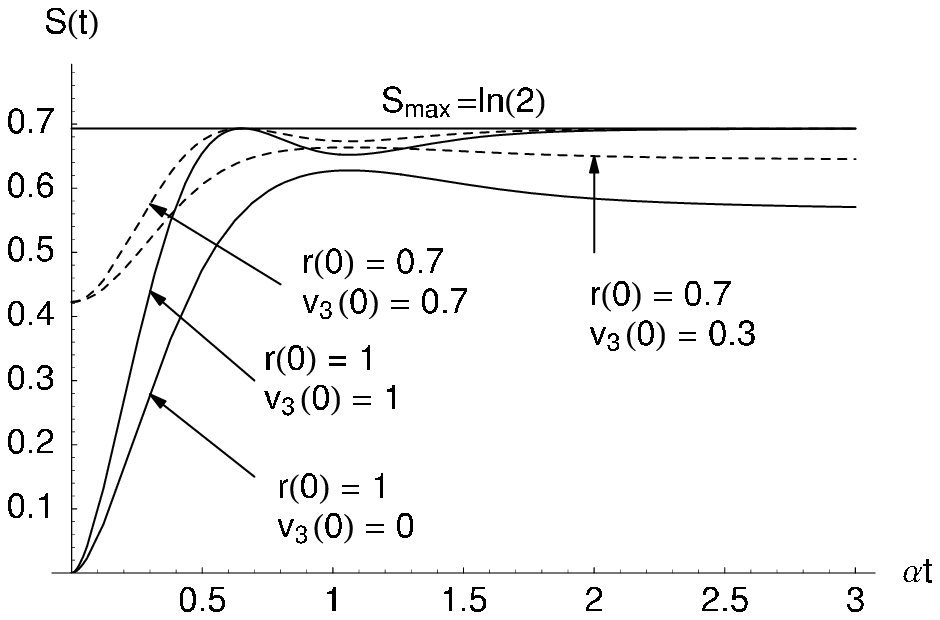}} \par}

\caption{\label{entro}Von Neumann entropy \protect\( S(t)\protect
\) of the reduced system for different initial conditions in the
limit \protect\( N\rightarrow \infty \protect \). \protect\(
S_{max}\equiv \ln 2\protect \) is the maximal entropy for a qubit,
representing a completely mixed state.}
\end{figure}

By contrast to \( \textrm{erf}(x) \), the imaginary error
function is not bounded. However, \( g(t) \) is bounded, and in
the limit \( t\rightarrow \infty  \) we have \( g(t)\rightarrow
-\frac{1}{2} \). Thus, in this limit the system is described by
the stationary density matrix
\begin{equation}
 \lim _{t\rightarrow
 \infty }\lim _{N\rightarrow \infty }\rho _{S}(t)=\left(
 \begin{array}{cc} \frac{1}{2} & \frac{v_{-}(0)}{2}\\
 \frac{v_{+}(0)}{2} & \frac{1}{2}
 \end{array}\right).
\end{equation}
The 3-component of the Bloch vector relaxes to zero, while
the off-diagonal elements of the density matrix show partial
decoherence, i.e. they assume half of their original values.
This behavior is also reflected in the von Neumann entropy of the
reduced system. Its dynamics depends on the initial entropy
parametrized by \( r(0) \) and on \( v_{3}(0) \), which is obvious
because the entropy is a scalar quantity and the system is
invariant under rotations around the \( z \)-axis. Figure
\ref{entro} shows the entropy as a function of time for different
initial conditions.

\section{Approximation techniques\label{approximations}}

In this section we will apply different approximation
techniques to the spin star model introduced and discussed in the
previous section. Due to the simplicity of this model we can not
only integrate exactly the reduced system dynamics, but also
construct explicitly the various master equations for the density
matrix of the central spin and analyze and compare their
perturbation expansions. In the following discussion we will stick
to the Bloch vector notation. Each of the master equations
obtained can easily be transformed into an equation involving
Lindblad superoperators [see Eqs.~(\ref{supops}), (\ref{supops2})] using the
translation rules
\begin{eqnarray}
 v_{3}\sigma _{3} &=& \left\{
 \frac{1}{2}\mathcal{S}_{3}-\mathcal{S}_{+}-\mathcal{S}_{-}\right\}
 \rho _{S}, \\
 v_{+}\sigma _{-}+v_{-}\sigma _{+} &=& -\frac{1}{2}\mathcal{S}_{3}\rho _{S}.
\end{eqnarray}

\subsection{Second order approximations\label{second}}

The second order approximation of the master equation for the
reduced system is usually obtained within the Born approximation
\cite{HBFP02b}. It is equivalent to the second order of the
Nakajima-Zwanzig projection operator technique, which will be
discussed systematically in Sec.~\ref{higher}. In our model the
Born approximation leads to the master equation
\begin{eqnarray}
 \lefteqn {\dot{\rho }_{S}(t)} & & \nonumber \\
 &=& -\int ^{t}_{0}ds\, \textrm{tr}_{B}
 \left\{ \left[ H,\left[ H,\rho _{S}(s)\otimes \rho _{B}(0)\right]
 \right] \right\} \nonumber \\
 &=& -8\alpha ^{2}Q_{1}\int ^{t}_{0}ds\left(
 v_{+}(s)\sigma _{-}+v_{-}(s)\sigma _{+}+v_{3}(s)\sigma_{3}\right),
 \qquad \label{integro}
\end{eqnarray}
where the bath correlation function is found to be
\begin{eqnarray}
 Q_{1} &=& \frac{1}{2^{N}} \textrm{tr}_{B}\left\{
 J_{+}J_{-} \right\} \label{q_1}
 \nonumber \\
 &=& \frac{1}{2^{N}} \textrm{tr}_{B}\left\{
 \sum _{i,j}\sigma ^{(i)}_{+}\sigma ^{(j)}_{-}\right\}
 \nonumber \\
 &=& \frac{1}{2^{N}} \textrm{tr}_{B}\left\{
 \sum _{i}\sigma ^{(i)}_{+}\sigma ^{(i)}_{-}\right\}
 \nonumber \\
 &=& \frac{N}{2}.
\end{eqnarray}
It is important to notice that \( Q_{1} \), as well as all
other bath correlation functions are independent of time. This is
to be contrasted to those situation in which the bath correlation
functions decay rapidly and which therefore allow the derivation
of a Markovian master equation. The time-independence of the
correlation functions is the main reason for the non-Markovian
behavior of the spin bath model.

The integro-differential equation (\ref{integro}) can easily
be solved by a Laplace transformation with the solution
\begin{eqnarray}
 \frac{v_{\pm }(t)}{v_{\pm }(0)} & = & \cos
 \left( 2\sqrt{N}\alpha t\right) ,\\ \frac{v_{3}(t)}{v_{3}(0)}
 &=& \cos \left( 2\sqrt{2N}\alpha t\right) .
\end{eqnarray}
In many physical applications the integration of the
integro-differential equation is much more complicated and one
tries to approximate the dynamics through a master
equation which is local in time. To this end, the terms \( v_{\pm
}(s) \) and \( v_{3}(s) \) under the integral in
Eq.~(\ref{integro}) are replaced by \( v_{\pm }(t) \) and \(
v_{3}(t) \), respectively. We thus arrive at the time-local master
equation
\begin{eqnarray}
 \lefteqn {\frac{d}{dt}\rho _{S}(t)} &  & \nonumber \\
 & = & -4N\alpha ^{2}\int ^{t}_{0}ds\left( v_{+}(t)\sigma _{-}
 +v_{-}(t)\sigma _{+}+v_{3}(t)\sigma _{3}\right) \qquad \nonumber \\
 & = & -4Nt\alpha ^{2}\left( v_{+}(t)\sigma _{-}+v_{-}(t)\sigma _{+}
 +v_{3}(t)\sigma _{3}\right), \qquad \label{redfield}
\end{eqnarray}
which is sometimes referred to as Redfield equation. Also
this master equation is easily solved to give the expressions
\begin{eqnarray}
 \frac{v_{\pm }(t)}{v_{\pm }(0)} & = & \exp (-2N\alpha ^{2}t^{2}),
 \label{tcl2a} \\
 \frac{v_{3}(t)}{v_{3}(0)} & = & \exp (-4N\alpha ^{2}t^{2}).
 \label{tcl2b}
\end{eqnarray}
The Redfield equation is equivalent to the second order of the
time-convolutionless projection operator technique, which will also be
discussed in detail in Sec.~\ref{higher}.

In order to obtain, finally, a Markovian master equation,
i.e. a time-local equation involving a time independent generator,
one pushes the upper limit of the integral in Eq.~(\ref{redfield})
to infinity, that is one studies the limit $t\rightarrow\infty$ of
the master equation. This limit leads to the Born-Markov
approximation of the reduced dynamics. In the present model,
however, it is not possible to perform this approximation because
the integrand does not vanish for large \( t \). Thus, the
Born-Markov limit does not exist for the spin bath model
investigated here and the description of relaxation and
decoherence processes requires the usage of non-Markovian
methods.

\subsection{Higher order approximations\label{higher}}

A systematic approach to obtain approximate non-Markovian
master equation in any desired order is provided by the projection
operator techniques. We define a projection superoperator
${\mathcal{P}}$ through the relation
\begin{equation}
 {\mathcal{P}}\rho = \textrm{tr}_{B}\{\rho\} \otimes \rho _{B}
\end{equation}
with the reference state \( \rho _{B}\equiv \rho _{B}(0) \) and
introduce the notation
\begin{equation} \left\langle
 \mathcal{X}\right\rangle \equiv \mathcal{PXP}
\end{equation}
for any superoperator \( \mathcal{X} \). Note that the
{}``moments'' \( \left\langle \mathcal{X}^{n}\right\rangle  \) are
operators in the total Hilbert space \( \mathcal{H}_{S}\otimes
\mathcal{H}_{B} \) of the combined system.

There are two main projection operator methods: The
Nakajima-Zwanzig (NZ) technique and the
time-convolutionless (TCL) technique. In our model, the
initial conditions factorize. The NZ and the TCL method
therefore lead to relatively simple, homogeneous equations of
motion. The NZ master equation is an integro-differential
equation for the reduced density matrix with a memory
$\mathcal{N}(t,\tau )$, which takes the form
\begin{equation}
 \dot{\rho }_{S}(t)\otimes \rho _{B} = \int^{t}_{0}d\tau
 \mathcal{N}(t,\tau )\rho _{S}(\tau )\otimes \rho_{B},
\end{equation}
while the TCL master equation is a time-local equation of
motion with a time-dependent generator $\mathcal{K}(t)$, which
reads
\begin{equation}
 \dot{\rho }_{S}(t)\otimes \rho _{B}
 = \mathcal{K}(t)\rho _{S}(t)\otimes \rho _{B}.
\end{equation}
Both the NZ and the TCL master equation can of course be
expanded with respect to the coupling strength $\alpha$. Since the
interaction Hamiltonian is time independent, this expansion
yields
\begin{equation}
 \int ^{t}_{0}d\tau \mathcal{N}(t,\tau )\rho _{S}(\tau )
 = \sum _{n=1}^{\infty }\alpha ^{n}\mathcal{I}_{n}(t,\tau )
 \left\langle \mathcal{L}^{n}\right\rangle _{pc}\rho _{S}(\tau )
\end{equation}
for the NZ master equation, and
\begin{equation}
 \mathcal{K}(t) = \sum ^{\infty }_{n=1}\alpha ^{n}
 \frac{t^{n-1}}{(n-1)!}\left\langle \mathcal{L}^{n}\right\rangle _{oc}
\end{equation}
for the TCL master equation, where we have introduced the
integral operator
\begin{equation}
 \begin{array}{c}
 \mathcal{I}_{n}(t,\tau ) \equiv
 \int ^{t}_{0}dt_{1}\int ^{t_{1}}_{0}dt_{2}\cdots
 \int ^{t_{n-3}}_{0}dt_{n-2}\int ^{t_{n-2}}_{0}d\tau
 \end{array}
\end{equation}
for the ease of notion. The symbol \( \left\langle
\mathcal{L}^{n}\right\rangle _{pc} \) denotes the partial
cumulants and \( \left\langle \mathcal{L}^{n}\right\rangle _{oc}
\) the ordered cumulants of order $n$. Their definitions
can be found in Refs. \onlinecite{FSTA80,vK74,NvK74,AR72}. In our model we have
\[
 \left\langle \mathcal{L}^{2n+1}\right\rangle_{pc}
 = \left\langle \mathcal{L}^{2n+1}\right\rangle _{oc}=0
\]
and
\begin{eqnarray*}
 \left\langle \mathcal{L}^{2}\right\rangle_{pc} & = & \left\langle
 \mathcal{L}^{2}\right\rangle, \\ \left\langle
 \mathcal{L}^{2}\right\rangle _{oc} & = & \left\langle
 \mathcal{L}^{2}\right\rangle, \\ \left\langle
 \mathcal{L}^{4}\right\rangle _{pc} & = & \left\langle
 \mathcal{L}^{4}\right\rangle -\left\langle
 \mathcal{L}^{2}\right\rangle ^{2}, \\ \left\langle
 \mathcal{L}^{4}\right\rangle _{oc} & = & \left\langle
 \mathcal{L}^{4}\right\rangle -3\left\langle
 \mathcal{L}^{2}\right\rangle ^{2}, \\ \left\langle
 \mathcal{L}^{6}\right\rangle _{pc} & = & \left\langle
 \mathcal{L}^{6}\right\rangle -2\left\langle
 \mathcal{L}^{2}\right\rangle \left\langle
 \mathcal{L}^{4}\right\rangle +3\left\langle
 \mathcal{L}^{2}\right\rangle ^{3}, \\ \left\langle
 \mathcal{L}^{6}\right\rangle _{oc} & = & \left\langle
 \mathcal{L}^{6}\right\rangle -15\left\langle
 \mathcal{L}^{2}\right\rangle \left\langle
 \mathcal{L}^{4}\right\rangle +30\left\langle
 \mathcal{L}^{2}\right\rangle ^{3}, \\
 & \cdots  & \nonumber
\end{eqnarray*}

In the time independent case the ordered cumulants are just
the ordinary cumulants know from classical statistics. To calculate
theses functions one can again use Eq. (\ref{L^2m_rho}). The
functions \( Q_{k} \) and \( R^{k-l}_{l} \) are real polynomials
in \( N \) of order \( k \). A method of determining these
polynomials is sketched in the appendix.

If we express the resulting master equations in terms of \( v_{\pm
}(t) \) and \( v_{3}(t) \), we get for the TCL technique
\begin{eqnarray}
 \textrm{TCL}:\; \dot{v}_{\pm }(t) & = &
 \left( \sum _{n=1}^{\infty }\alpha^{2n}
 \frac{s_{2n}t^{2n-1}}{(2n-1)!}\right) v_{\pm}(t),
 \label{TCLeq1} \\
 \dot{v}_{3}(t) & = & \left( \sum_{n=1}^{\infty }\alpha ^{2n}
 \frac{2q_{2n}t^{2n-1}}{(2n-1)!}\right) v_{3}(t),
 \label{TCLeq2}
\end{eqnarray}
and for the NZ method
\begin{eqnarray}
 \textrm{NZ}:\;\dot{v}_{\pm }(t) & = & \left( \sum _{n=1}^{\infty }
 \alpha^{2n}\tilde{s}_{2n}\mathcal{I}_{n}(t,\tau )\right) v_{\pm }(\tau),
 \label{NZeq1} \\
 \dot{v}_{3}(t) & = & \left( \sum _{n=1}^{\infty}\alpha ^{2n}
 2\tilde{q}_{2n}\mathcal{I}_{n}(t,\tau )\right)v_{3}(\tau ).
 \label{NZeq2}
\end{eqnarray}
The quantities $s_{2n}$, $\tilde{s}_{2n}$, $q_{2n}$ and
$\tilde{q}_{2n}$ represent real polynomials in \( N \) of the
order \( n \). For example, we have
\begin{eqnarray*}
 q_{2} & = & -4N,\\
 q_{4} & = & -32N^{2},\\
 q_{6} & = & -1024N+1536N^{2}-1536N^{3},\\
 & \cdots  & \nonumber \\
 \tilde{q}_{2} & = & -4N,\\
 \tilde{q}_{4} & = & 32N^{2},\\
 \tilde{q}_{6} & = & -1024N+1536N^{2}-1280N^{3},\\
 & \cdots  & \nonumber \\
 s_{2} & = & -4N,\\
 s_{4} & = & -48N+16N^{2},\\
 s_{6} & = & -1024N-384N^{2}+384N^{3},\\
 & \cdots  & \nonumber \\
 \tilde{s}_{2} & = & -4N,\\
 \tilde{s}_{4} & = & -48N+48N^{2},\\
 \tilde{s}_{6} & = & -1024N+2112N^{2}-1216N^{3},\\
 & \cdots  & \, .\nonumber
\end{eqnarray*}

The $(2n)$th-order approximation of the master equations
(denoted by TCL$2n$ and NZ$2n$, respectively) is obtained by
truncating the sums in Eqs.~(\ref{TCLeq1}), (\ref{TCLeq2}) and in
Eqs.~(\ref{NZeq1}), (\ref{NZeq2}) after the $n$th term. In the TCL
case, the resulting ordinary differential equations can be
integrated very easily. The equation of motion of the NZ method
can be solved with the help of a Laplace transformation. However,
it may be very involved to carry out the inverse transformation
for higher orders. For example, the solution of the twelfth order
of the NZ equation as obtained by standard computer algebra tools
 is filling some hundred pages, whereas the
solution of the TCL equation can be written in a single line.

\begin{figure}[htba]
{\centering
\resizebox*{0.47\textwidth}{!}{\includegraphics{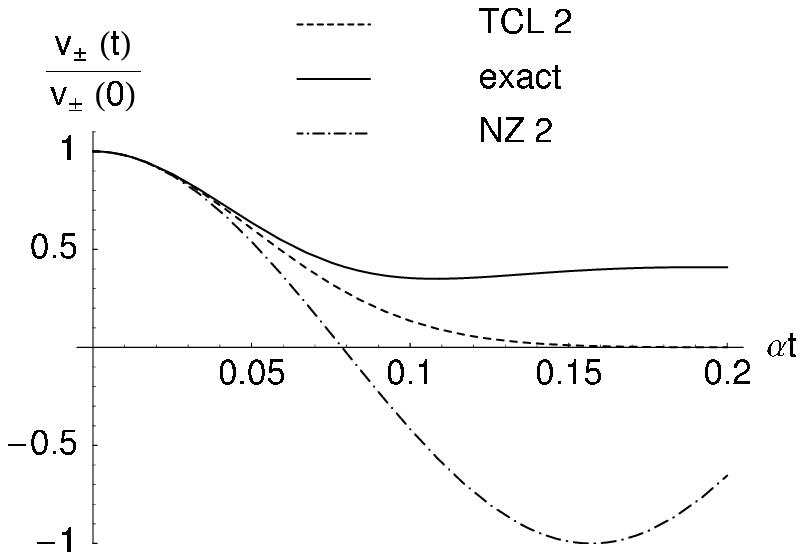}}
\resizebox*{0.47\textwidth}{!}{\includegraphics{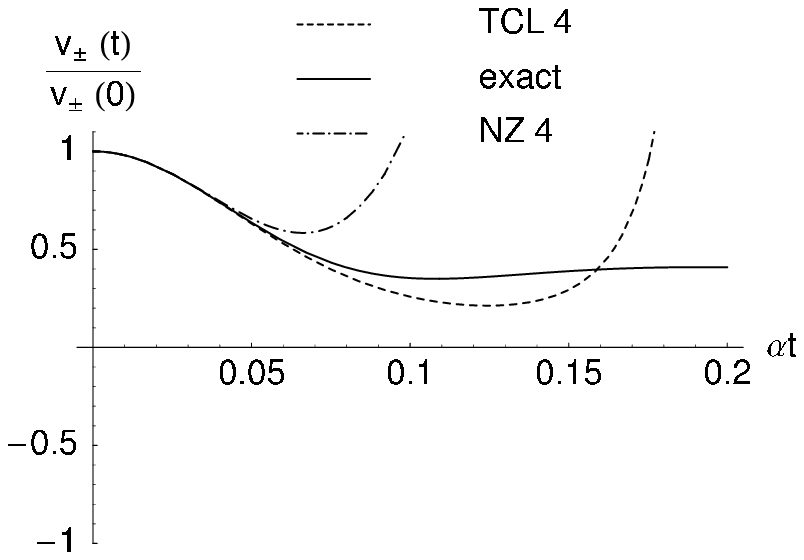}}
\par}

\caption{\label{plots_vergleich_pm_n}Comparison of the TCL
and the NZ technique with the exact solution. The plot shows the
approximations to second and fourth order in \protect\( \alpha
\protect \) and the exact solution of \protect\( v_{\pm
}(t)\protect \) [see Eqs.~(\ref{bloch1}) and (\ref{F12})] for a
bath of \protect\( 100\protect \) spins. }
\end{figure}

\begin{figure}[htba]
{\centering
\resizebox*{0.47\textwidth}{!}{\includegraphics{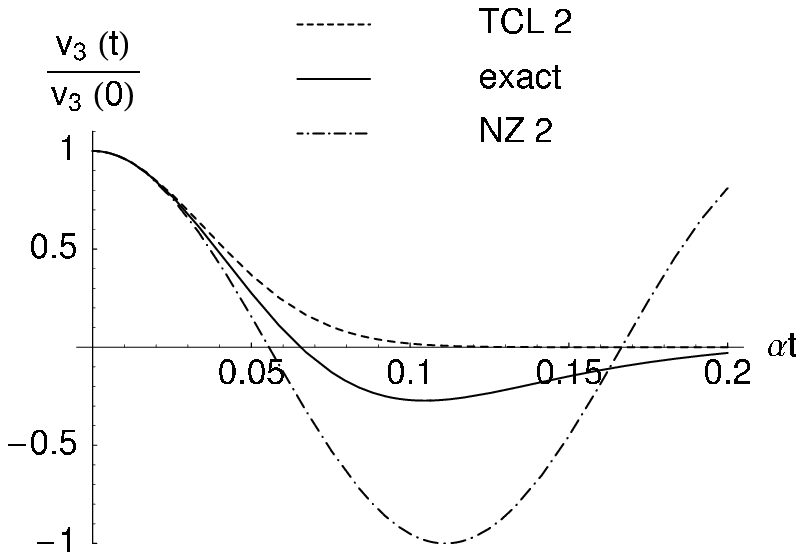}}
\resizebox*{0.47\textwidth}{!}{\includegraphics{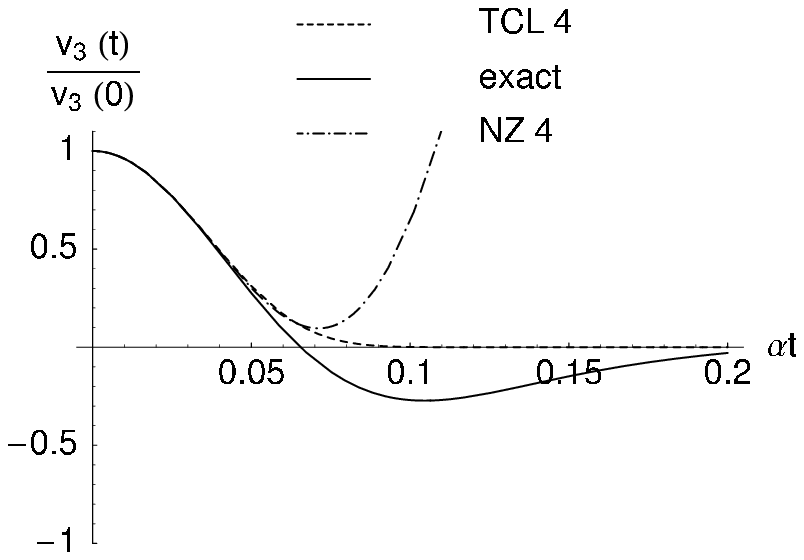}}
\par}

\caption{\label{plots_vergleich_3_n}Comparison of the TCL and
the NZ technique with the exact solution. The plot shows the
second and the fourth order approximations as well as the exact
solution of \protect\( v_{3}(t)\protect \) [see
Eqs.~(\ref{bloch2}) and (\ref{F3})] for a bath of \protect\(
100\protect \) spins. }
\end{figure}

The solutions of the master equations in second and fourth
order are plotted in Figs.~\ref{plots_vergleich_pm_n} and
\ref{plots_vergleich_3_n}, together with the exact solutions. We
observe that both methods lead to a good approximation of the
short-time behavior of the components of the Bloch vector. We
further see that the TCL technique is not only easier to solve,
but also provides a better approximation of the dynamics within a
given order.

\begin{figure}[htba]
{\centering
\resizebox*{0.47\textwidth}{!}{\includegraphics{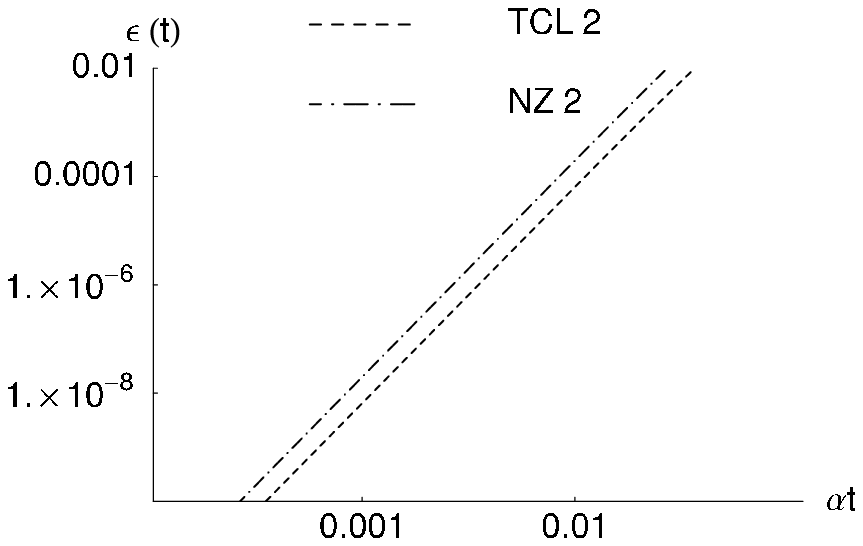}}\par}

\caption{\label{error}Error of TCL2 and NZ2: The plot shows
the deviation \protect\(
\epsilon(t)\equiv\left|v_{\pm}(t)-v^{\textrm{approx}}_{\pm
}(t)\right| \protect \) of the exact solution $v_{\pm}(t)$ from
the approximate solution $v^{\textrm{approx}}_{\pm }(t)$ for small
\protect\( \alpha t\protect \).}
\end{figure}

Since the TCL and the NZ method lead to expansions of the
equations of motion and not of their solutions, the solutions of
the truncated equations may contain terms of arbitrary order in
the coupling strength. For example, even though TCL2 is a second
order approximation, the corresponding solution given by
Eqs.~(\ref{tcl2a}) and (\ref{tcl2b}) contains infinitely many
orders. Of course, the expansion of the exact solution coincides
with the expansion of the approximations obtained with TCL$2n$ or
NZ$2n$ within the $(2n)$th order. The error of TCL2 or NZ2, for
example, is therefore a term of order $\alpha ^{4}$, as is
illustrated in Fig.~\ref{error}.

Concerning the long-time behavior, both the TCL and the NZ
method may lead to very bad approximations. For example, in the
fourth order approximation of \( v_{\pm }(t) \) (see
Fig.~\ref{plots_vergleich_pm_n}) the TCL as well as the NZ
solution leave the Bloch sphere, i.e. for times larger than some
critical time these solutions do not represent true density
matrices anymore.

\begin{figure}[htba]
{\centering \resizebox*{0.47\textwidth}{!}{\includegraphics{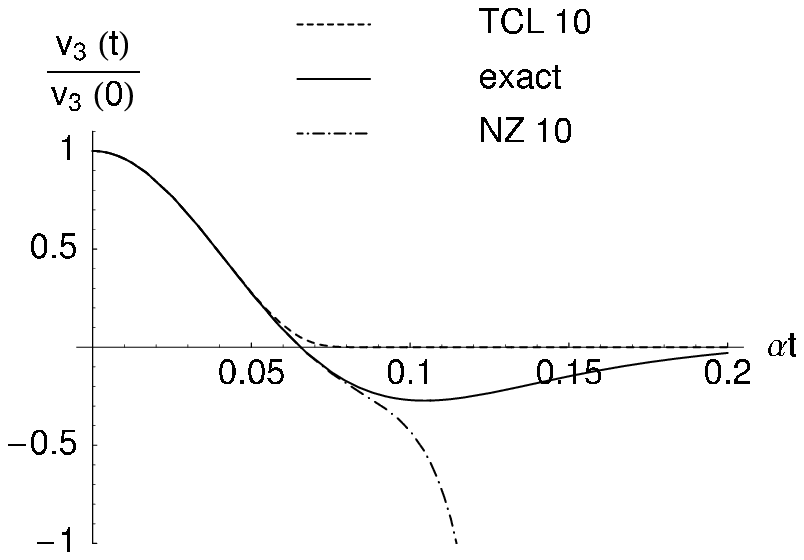}}
\resizebox*{0.47\textwidth}{!}{\includegraphics{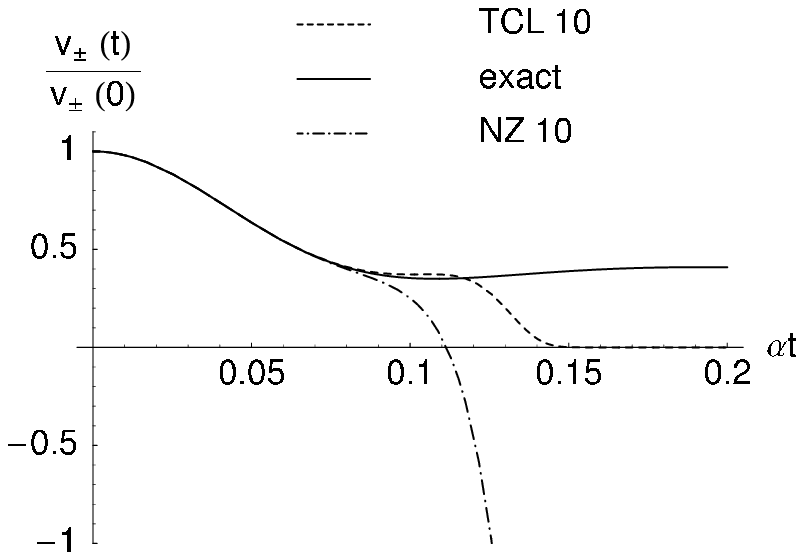}} \par}

\caption{\label{plots_vergleich2_n}The TCL and the NZ
approximation of the components of the Bloch vector in tenth
order for a bath of \protect\( 100\protect \) spins. }
\end{figure}

If we look at higher orders, the NZ method is seen to be
better than the TCL method as far as the 3-component of
the Bloch vector is concerned. An example is shown in
Fig.~\ref{plots_vergleich2_n}, where we plot the tenth order
approximations. We observe that the solution \( v_{3}(t) \) of the
TCL equation (\ref{TCLeq2}) is always greater than zero. This fact is obviously
connected to the structure of this equation, which takes the form 
$\dot{v}_{3}(t)=\mathcal{K}_{3}(t)v_{3}(t)$ with a real function $\mathcal{K}_{3}(t)$.
If the $3$-component $v_{3}$ of the Bloch vector vanishes at the time $t=t_{0}$, then a 
time-local equation of motion of this form can only be fulfilled if $\dot{v}_{3}(t_{0})$
is also zero. In our case, however, the exact solution passes zero with a non-vanishing
time derivative. It is a well-known fact \cite{HBFP02b} that the perturbation expansion of
the TCL generator only exists, in general, for short and intermediate
times and/or coupling
strengths. This is reflected in the fact that the superoperator on
the right-hand side of Eq.~(\ref{supop_rho_s_von_t}) cannot be
inverted for all times, i.e. it is not always possible to express
\( v_{3}(0) \) in terms of \( v_{3}(t) \). A similar situation
also occurs in open systems interacting with a bosonic reservoir,
e.~g., in the damped Jaynes Cummings model which describes the
interaction of a qubit with a bosonic reservoir at zero temperature.
The NZ technique does not lead to such problems. However, since
the components $v_{\pm}(t)$ do not vanish, the corresponding
high-order TCL approximation is still more accurate than the NZ
approximation, as may be seen from
Fig.~\ref{plots_vergleich2_n}.

\section{Conclusion\label{CONCLU}}

With the help of a simple analytically solvable model of a
spin star system, we have discussed the performance
of projection operator techniques for the dynamics of open systems
and the resulting perturbation expansions of the equations of
motion. The model consists of a central spin surrounded by a bath
of spins interacting with the central spin through a Heisenberg \(
XX \) coupling, and shows complete relaxation and partial
decoherence in the limit of an infinite number of bath spins. Due
to its high symmetry the model allows a direct comparison of the
Nakajima-Zwanzig (NZ) and of the time-convolutionless (TCL)
projection operator methods with the exact solution in analytical
terms.

While the Born-Markov limit of the equation of motion does not
exist in the model, the dynamics of the central spin exhibits a
pronounced non-Markovian behavior. It has been demonstrated that
both the NZ and the TCL techniques provide good approximations of
the short-time dynamics. In practical applications the TCL method
is usually to be preferred since it leads to time-local equations
of motion in any desired order with a much easier mathematical
structure, whose integration is much simpler than that of the
non-local equations of the NZ technique.

It should be kept in mind, however, that the expansion based on
the TCL method converges, in general, only for short and
intermediate interaction times. For large times the perturbation
expansion may break down, which has been illustrated in our model
to be connected to zeros of the components of the Bloch vector. It
turns out that the NZ equation of motion yields a better
approximation of the exact dynamics in this regime.

In view of the heuristic approach to the Born and to the Redfield
equation (see Sec.~\ref{second}) it is sometimes conjectured that
a non-local equation of motion should be generally better than a
time-local one. The results of Sec.~\ref{higher} show that this
conjecture is, in general, not true. The fact that in a given
order the time-local TCL equation is at least as
good (and much simpler to deal with), if not better than the
non-local NZ equation has also been observed in other specific
system-reservoir models \cite{HBFP02b}, and has been confirmed by general
mathematical arguments \cite{AR03}. However, care must be taken
when applying a certain projection operator method to a specific
model: The quality of the corresponding perturbation expansion
of the equation of motion may strongly depend on the specific
properties of the model, e.~g., the interaction Hamiltonian, the interaction time,
the environmental state and the spectral density.

For example, in our particular model the TCL expansion to fourth
order turns out to be more accurate than the fourth order NZ
expansion. However, there is no reason why TCL should be generally
better then NZ. To clarify further this point we consider the
Taylor series of the 3-component of the Bloch vector:
\begin{equation}
\label{expansion_v3}
 v_{3}(t) = a_{0}+a_{2}\left(\alpha t\right)^{2}
 + a_{4}\left( \alpha t \right)^{4}
 +\mathcal{O}\left((\alpha t)^{6}\right).
\end{equation}
The corresponding expansion obtained from TCL2 is given by
\begin{equation}
 v_{3}(t) = a_{0}+a_{2}\left( \alpha t \right)^{2}
 +\frac{a^{2}_{2}}{2a_{0}}\left( \alpha t \right)^{4}
 +\mathcal{O}\left((\alpha t)^{6}\right),
\end{equation}
while NZ2 gives the expansion
\begin{equation}
 v_{3}(t) = a_{0}+a_{2}\left(\alpha t \right)^{2}
 +\frac{a^{2}_{2}}{6a_{0}}\left(\alpha t \right)^{4}
 +\mathcal{O}\left( (\alpha t)^{6}\right).
\end{equation}
In our model the exact coefficients of the expansion (\ref{expansion_v3}) are found to be
\begin{equation}
 a_{0} = 1, \qquad a_{2} = -4N, \qquad a_{4} = \frac{16}{3}N^{2}.
\end{equation}
Of course, the second order coefficient $a_2$ is the same in all
expansions, while in general neither TCL2 nor NZ2 reproduce correctly the
fourth order coefficient $a_4$. However, in our model
it turns out that the TCL2 approximation is more accurate because
the fourth order coefficient
\begin{equation}
 \frac{a^{2}_{2}}{2a_{0}}=\frac{16}{2}N^{2}
\end{equation}
found from the solution of the TCL equation is closer to the
correct fourth order coefficient \( a_{4} \) than the
corresponding coefficient
\begin{equation}
 \frac{a^{2}_{2}}{6a_{0}}=\frac{16}{6}N^{2}
\end{equation}
of the NZ equation (see Fig.~\ref{error}). Thus we see that it depends crucially on the
value of \( a_{4} \) whether TCL2 or NZ2 is better.

Choosing an appropriate interaction Hamiltonian and initial state,
one can easily construct examples where NZ2 is better than TCL2.
For example, if \( v_{3}(t) \) was a cosine function
$a_{0}\cos(\alpha t)$, then NZ2 would already give the exact
solution. On the other hand, if \( v_{3}(t) \) was a Gaussian
function $a_{0}\exp(-\alpha^2t^2)$, then TCL2 would reproduce the
exact solution because the higher cumulants of a Gaussian function
vanish.

The features discussed above should be taken into account in
applications of projection operator methods to specific open
systems. In the general case in which an analytical solution is
not known a careful analytical or numerical investigation of the
higher orders of the respective expansions is thus indispensable
to judge the quality of the TCL or the NZ method, the influence of
initial correlations, or to estimate the timescale over which one
can trust the approximation obtained within a given order. 

\appendix
\section{Bath Correlation Functions\label{appendix1}}

In this appendix we outline how to calculate the bath correlation
functions
\begin{eqnarray}
 Q_{k} & \equiv & \frac{1}{2^{N}}\textrm{tr}_{B}\left\{
 \left( J_{+}J_{-}\right) ^{k}\right\}, \\
 R^{k-l}_{l} & \equiv  & \frac{1}{2^{N}}\textrm{tr}_{B}\left\{
 \left( J_{+}J_{-}\right) ^{k-l}\left( J_{-}J_{+}\right) ^{l}\right\} .
\end{eqnarray}
The trace can be computed in the eigenbasis of \( J_{3} \) and \(
\mathbf{J}^{2} \) (see Sec.~\ref{reduceddyn}) yielding a sum of
polynomials in \( j \) and \( m \). However, it turns out that it
is easier to use the computational basis of the spin bath
consisting of the states
\begin{equation}
 \ket{s_{1}}\otimes \ket{s_2}\otimes \cdots \otimes \ket{s_N},
\end{equation}
where the $s_{i}$ take on the values $0$ or $1$ and
\begin{equation}
 \sigma^{(i)}_3\ket{s_{i}}=(-1)^{s_{i}}\ket{s_{i}}.
\end{equation}
With the help of these states the problem is reduced to a
combinatorial one. Since
\begin{equation}
 J_{\pm} = \sum_{i=1}^N \sigma _{\pm }^{(i)}
\end{equation}
we have
\begin{equation} \label{summe}
 \left( J_{+}J_{-}\right)^{k} =
 \sum_{i_{1},\ldots,i_{2k}}
 \sigma^{(i_{1})}_{+}\sigma ^{(i_{2})}_{-}\sigma^{(i_{3})}_{+}
 \sigma^{(i_{4})}_{-}
 \cdots \sigma^{(i_{2k-1})}_{+}\sigma ^{(i_{2k})}_{-},
\end{equation}
where the summation is taken over all possible combinations
of the indices $i_1,i_2,\ldots,i_{2k}$. Under the trace over the
bath we can sort these indices, without interchanging the
operators belonging to the same index, and calculate the partial
traces over the various bath spins separately.

Let us denote the partial trace over the Hilbert space of the
$i$th bath spin by $\textrm{tr}_i$. For example, we have for
$k=2$:
\begin{eqnarray}
 \lefteqn {\textrm{tr}_{B}\left\{
 \sigma ^{(1)}_{+}\sigma ^{(3)}_{-}\sigma ^{(4)}_{+}\sigma ^{(1)}_{-}\right\} }
 &  & \nonumber \\
 & = & \textrm{tr}_{1}\left\{ \sigma ^{(1)}_{+}\sigma ^{(1)}_{-}\right\}
 \textrm{tr}_{3}\left\{ \sigma ^{(3)}_{-}\right\} \textrm{tr}_{4}\left\{
 \sigma ^{(4)}_{+}\right\} 2^{N-3} \nonumber \\
 & = & 0,
\end{eqnarray}
since $\textrm{tr}_{i}\left\{ \sigma^{(i)}_{\pm}\right\}=0$.
Note that the factor \( 2^{N-3} \) appears due to \( (N-3) \)
factors of \( \textrm{tr}_{i}\left\{ I \right\} =2 \). These
factors arise from the partial traces of the unit matrices $I$ in
the spin spaces, which we did not write explicitly. As a further
example, we have for \( k=4 \):
\begin{eqnarray}
 \lefteqn {\textrm{tr}_{B}\left\{
 \sigma^{(1)}_{+}\sigma^{(3)}_{-}\sigma ^{(1)}_{+}\sigma ^{(1)}_{-}
 \sigma^{(3)}_{+}\sigma ^{(1)}_{-}\right\} } &  & \nonumber \\
 & = & \textrm{tr}_{1}\left\{
 \sigma^{(1)}_{+}\sigma ^{(1)}_{+}\sigma ^{(1)}_{-}\sigma^{(1)}_{-}\right\}
 \textrm{tr}_{3}\left\{ \sigma ^{(3)}_{-}\sigma ^{(3)}_{+}\right\} 2^{N-2}\quad
 \nonumber \\
 & = & 0,
\end{eqnarray}
because of \( \sigma_{\pm}^{(i)}\sigma_{\pm}^{(i)}=0 \). An
example of a non-vanishing term is given by
\begin{eqnarray}
 \lefteqn {\textrm{tr}_{B}\left\{
 \sigma ^{(1)}_{+}\sigma ^{(2)}_{-}\sigma ^{(2)}_{+}\sigma ^{(1)}_{-}\right\} }
 &  & \nonumber \\
 & = & \textrm{tr}_{1}\left\{ \sigma ^{(1)}_{+}\sigma ^{(1)}_{-}\right\}
 \textrm{tr}_{2}\left\{ \sigma^{(2)}_{-}\sigma^{(2)}_{+}\right\} 2^{N-2}
 \nonumber \\
 & = & 2^{N-2}, \quad
\end{eqnarray}
where we have used that
${\mathrm{tr}}_{i}\left\{\sigma^{(i)}_{\mp}\sigma^{(i)}_{\pm}\right\}=1$.

In view of these considerations we are now left with the
combinatorial problem of determining all nonzero summands for the
given values of $k$ and $l$. As an example, let us calculate
explicitly the correlation function \( Q_{2} \). From its
definition we have
\begin{eqnarray}
 Q_{2} & = & \frac{1}{2^{N}}\textrm{tr}_{B}
 \left\{ J_{+}J_{-}J_{+}J_{-}\right\} \nonumber \\
 & = & \frac{1}{2^{N}} \sum _{i_{1},i_{2},i_{3},i_{4}} \textrm{tr}_{B} \left\{
 \sigma ^{(i_{1})}_{+}\sigma ^{(i_{2})}_{-}\sigma^{(i_{3})}_{-}\sigma ^{(i_{4})}_{+}\right\}.\quad
\end{eqnarray}
The nonzero summands in this expression have the following
structure:
\begin{eqnarray*}
 \sigma ^{(i)}_{+}\sigma ^{(i)}_{-}\sigma ^{(i)}_{+}\sigma ^{(i)}_{-}
 & \rightarrow  & N\textrm{ possibilities,}\\
 \sigma ^{(i)}_{+}\sigma ^{(i)}_{-}\sigma ^{(j)}_{+}\sigma ^{(j)}_{-}
 & \rightarrow  & N(N-1)\textrm{ possibilities,}\\
 \sigma ^{(i)}_{+}\sigma ^{(j)}_{-}\sigma ^{(j)}_{+}\sigma ^{(i)}_{-}
 & \rightarrow  & N(N-1)\textrm{ possibilities,}
\end{eqnarray*}
where $i \neq j$ in the second and the third line. Collecting
these results we find
\begin{eqnarray}
 Q_{2} & = & \frac{1}{2^{N}}\left(
 N \cdot 2^{N-1} + 2\cdot N(N-1)\cdot 2^{N-2}\right) \quad
 \nonumber \\
 & = & \frac{N^{2}}{2}.
\end{eqnarray}
A similar procedure must be carried out to calculate \(
R^{k-l}_{l} \). We state some results:
\begin{eqnarray*}
 Q_{3} & = & \frac{1}{2}N-\frac{3}{4}N^{2}+\frac{3}{4}N^{3},\\
 Q_{4} & = & -2N+5N^{2}-4N^{3}+\frac{3}{2}N^{4},\\
 & \cdots  &  \\
 R_{1}^{1} & = & -\frac{1}{2}N+\frac{1}{2}N^{2},\\
 R^{1}_{2} & = & \frac{1}{2}N-\frac{5}{4}N^{2}+\frac{3}{4}N^{3},\\
 R^{1}_{3} & = & -\frac{5}{2}N+\frac{23}{4}N^{2}-\frac{19}{4}N^{3}+\frac{3}{2}N^{4},\\
 & \cdots  & \, .
\end{eqnarray*}
It should be clear that the above method of determining the
correlation functions is easily translated into a numerical code
from which one obtains the $Q_k$ and the $R_l^{k-l}$ in any
desired order.

For $N \geq k$, the term of leading order in $N$ of the
polynomials $Q_k$ and $R_l^{k-l}$ is represented by the summands
with a maximal number of $k$ different indices, because these
terms have the largest combinatorial weight. After sorting the
spin operators, these terms will have the following form,
\begin{equation} \label{GENFORM}
 \sigma ^{(i_{1})}_{+}\sigma ^{(i_{1})}_{-}
 \sigma^{(i_{2})}_{+}\sigma ^{(i_{2})}_{-}\cdots
 \sigma ^{(i_{k})}_{+}\sigma ^{(i_{k})}_{-}.
\end{equation}
There are $\binom{N}{k}$ different ways of assigning the indices
$i_1,i_2,\ldots,i_k$ to this term. For a fixed set of indices,
there are $k! \cdot k!$ different terms in the sum (\ref{summe})
which lead to the sorted expression (\ref{GENFORM}), corresponding
to a permutation of the labels of all $\sigma_+$ operators and of
all $\sigma_-$ operators. The trace of the expression
(\ref{GENFORM}) yields $2^{N-k}$. Hence, the term of leading order
of the polynomial $Q_k$ is found to be
\begin{equation}
 2^{-N} \binom{N}{k} k! \cdot k! \, 2^{N-k}\approx \frac{N^{k}k!}{2^{k}}.
\end{equation}
A similar proof holds for \( R^{k-l}_{l} \). Thus, we have
for $N\rightarrow\infty$ and $k$ fixed:
\begin{equation} \label{ninfty}
 Q_{k}\approx R^{k-l}_{l}\approx \frac{k!}{2^{k}}N^{k},
\end{equation}
which has been used in Sec.~\ref{limit}.

\end{document}